# A protostellar system fed by a streamer of 10,500 au length

Jaime E. Pineda[1], Dominique Segura-Cox[1], Paola Caselli[1], Nichol Cunningham[2], Bo Zhao[1], Anika Schmiedeke[1], Maria José Maureira[1], Roberto Neri[2]

[1]Max-Planck-Institut für extraterrestrische Physik (MPE), Gießenbachstr. 1, D-85748 Garching, Germany
[2]Institut de Radioastronomie Millimétrique (IRAM), 300 rue de la Piscine, F-38406, Saint-Martin d'Hères, France

**Binary formation is an important aspect of star formation. One possible route for close-in binary formation is disk fragmentation[1,2,3]. Recent observations show small scale asymmetries (<300 au) around young protostars[2,4], although not always resolving the circumbinary disk, are linked to disk phenomena[5,6]. In later stages, resolved circumbinary disk observations[7] (<200 au) show similar asymmetries, suggesting the origin of the asymmetries arises from binary-disk interactions[8,9,10]. We observed one of the youngest systems to study the connection between disk and dense core. We find for the first time a bright and clear streamer in chemically fresh material (Carbon-chain species) that originates from outside the dense core (>10,500 au). This material connects the outer dense core with the region where asymmetries arise near disk scales. This new structure type, 10x larger than those seen near disk scales, suggests a different interpretation of previous observations: large-scale accretion flows funnel material down to disk scales. These results reveal the under-appreciated importance of the local environment on the formation and evolution of disks in early systems[13,14] and a possible initial condition for the formation of annular features in young disks[15,16].**

Per-emb-2 (IRAS 03292+3039) is a protostellar system at a distance of 300 parsecs[17] in the Perseus molecular cloud. It hosts a young forming close binary (<20 au) stellar system. CO (2-1) observations show a clear and well-collimated outflow[18]. The chemical abundances close to the young system are inconsistent with the current luminosity and provide strong evidence for a recent accretion burst. Independent studies need an accretion rate increase of a factor of 10x compared to the current one ($\dot{M}_{burst} = 7 \times 10^{-6}$ $M_{Sun}$ yr$^{-1}$)[19,20]. Thus, this protostellar system is a great target to study the role of the environment in accretion variability. Moreover, higher angular resolution observations reveal feather-like features near disk scales (<300 au). These and the velocity gradient perpendicular to the outflow orientation are claimed as clear evidence of a disk undergoing gravitational instability (GI)[4], but the detailed gas kinematic information showing clear Keplerian rotation is lacking to confirm this scenario.

The general picture of star formation usually involves an axisymmetric envelope (or dense core)[10,22]. This envelope flattens at small radii due to the material's angular momentum and/or magnetic field. It is during this process that a disk can form. Numerical simulations of disk formation usually focus on isolated systems[5,23], while those that track disk and star formation in a molecular cloud reveal asymmetric accretion flows[24,25,26]. These flows would change the accretion rates in protostars compared to an axisymmetric infall. No previous observations have revealed the presence of these large accretion flows. Recently, ALMA observations revealed evidence for small (less than 1000 au) accretion streams originating from within the dense core[27,11,12,28]. We observed several molecular lines using the NOrthern Extended Millimeter Array (NOEMA), see Figure 1, to unveil a streamer of fresh material (where atomic carbon is not yet mainly converted in CO) from beyond 10,500 au down to the disk forming scales (<300 au).

Proximity on the sky does not imply that objects are physically related. But, if the line-of-sight velocities are similar, then they are likely physically associated. Moreover, the gas kinematics are crucial to determine the streamer's dynamical state. We derive the gas velocity and velocity dispersion along the streamer (see Figure 2 and Methods section). This velocity map shows a smooth velocity gradient, with the largest velocity difference at the streamer's end. The velocities near the central region agree with that of the protostellar system. And the velocity dispersions are narrow and consistent with subsonic turbulence, which reveals the streamer as unperturbed by the outflow. The measured velocity gradient is 11.5 km s$^{-1}$ pc$^{-1}$, which is within 60% of the pure free-fall value (see Methods section). This suggests that the gravitational pull of the central dense core (3.2 M$_{Sun}$)[46] dominates the streamer's dynamics. The corresponding free-fall timescale is 96 kyr. This timescale is comparable to the embedded star-formation phase timescale (100 kyr)[22]. We further model the streamer as a flow of gas with an initial rotation and with its dynamics dominated by the gravity of the central dense core. The analytic solutions[29] reproduce both the position in the plane of the sky and the measured velocity along the line of sight (see Figure 3 and Methods section). This confirms that the streamer is fully consistent with free-fall after considering all the projection effects. Thus, streamers could be present during the full embedded phase and detectable in more objects.

Thanks to NOEMA's new wide-band correlator we observe many lines simultaneously. In particular, we image two different transitions of the HC$_3$N, (10-9) and (8-7), and both trace the streamer. The ratio between those lines in the streamer is 0.48±0.02, after primary beam correction is applied. We compare this ratio to predictions from RADEX[30] (for temperatures of 10 and 12.25 K, see Method section and Extended Data Figure 1) and determine that the streamer's average H$_2$ density and HC$_3$N column density are (4±2)x10$^4$ cm$^{-3}$ and 2.0x10$^{13}$ cm$^{-2}$ over a 10,500 au region (marked by the polygon in Figure 3). The constraints on the HC$_3$N/H$_2$ ratio, based on dense cores measurements, imply a total streamer mass between 0.1 to 1 M$_{Sun}$ (see Method section for details). Even the lowest mass estimate is a substantial fraction of the total dense core mass (3.2 M$_{Sun}$). As a crosscheck, we use the streamer's H$_2$ density, HC$_3$N column density, and typical width of 4,500 au to estimate a [HC$_3$N/H$_2$]$_{Streamer}$ abundance of 7.4x10$^{-9}$. This derived abundance is comparable to the upper end of the HC$_3$N abundance range for dense cores. Therefore, we adopt a conservative estimate of the streamer mass of 0.1 M$_{Sun}$, and it will be the value used hereafter. This value highlights the relevance of a previously unseen accretion route.

We estimate the streamer infall rate as $\dot{M}_{Streamer} = M_{Streamer}/t_{ff}$ = 10$^{-6}$ M$_{Sun}$ yr$^{-1}$. This is the mean infall rate from streamer to disk scales, and it is comparable (within a factor of 3x) to the mass infall rate of protostellar envelopes estimated in other young objects[31]. The current accretion rate estimated from the bolometric luminosity is 7x10$^{-7}$ M$_{sun}$ yr$^{-1}$ [19,20], see Method section for details. Thus the streamer infall rate is comparable to the current protostellar accretion rate. So, the streamer could change the protostellar accretion by funneling extra material to the central region. In fact, this stellar system has undergone at least one accretion burst of at least a factor of 10x in the last 10,000 yrs[19,20]. This further suggests that accretion events from the outer envelope (>3,000 au)[32] could drive variability of young sources[33].

The material in the inner envelope (<3,000 au) of Per-emb-2 is well traced by the simultaneous observations of N$_2$H$^+$ and N$_2$D$^+$ (see Extended Data Figure 2). These tracers are the best probes of the chemically evolved cores[34]. But these species fail to trace the streamer. In fact, Carbon-chain molecules show the best images of the streamer. These Carbon-chain species are the best probes of the less chemically evolved cores[34], or "chemically fresh" material. The abundance of Carbon-chains in the streamer reveals that its material is chemically fresh in comparison to the inner core. From the streamer density the depletion timescale is ~20,000 yr, which is comparable to the free-fall time scale, but for the inner dense core the depletion timescale is ~600 yrs. This timescale difference helps to explain the chemical difference between streamer and core (see Extended Data Figure 3). The streamer must indeed bring chemically fresh material from beyond (or in the outer regions of) the dense core on a timescale that is short relative to the canonical ~100,000 yr duration of the Class 0 phase. This also explains why previous observations using chemically evolved probes missed the streamer[18,20], because those species do not preferentially trace the streamer area and are more smoothly distributed within the dense core. Moreover, the chemically fresh material in the streamer should affect the chemistry close to the disk by bringing Carbon-chain rich material inwards[35].

The clear identification of such a large reservoir of fresh material in almost free-fall is remarkable in itself. It reveals that new material might shape the morphology and kinematics of the gas in young stellar systems. In particular, it gives support to studies claiming that streamer-like features seen at disk scales are the result of strong envelope accretion[32]. Although some numerical simulations show similar streamers[24,25,26], it is unclear how frequent and for how long in the protostar evolution this process could occur. If these streamers are present in more evolved systems, then they might generate the initial over densities necessary to generate rings before the planet formation process is in full swing[15,16]. And even in later stages, when planet formation is ongoing, a streamer could deposit material in the inner disk as required to explain the differences between different meteorites[36]. Our results show that streamers could originate from outside (or in the outer regions of) the dense core (>10,000 au, larger than the previously seen 1,000 au scales), and therefore non-axisymmetric accretion flows might be crucial ingredients in the star- and disk-formation process. Additional high-angular resolution observations of young protostars in more regions will determine the frequency of these streamers.

**References**

1. Kratter, K.M., Matzner, C.D., Krumholz, M.R. & Klein, R.I. *Astrophys. J.* **708**, 1585-1597 (2010).
2. Tobin, J.J. et al. A triple protostar system formed via fragmentation of a gravitationally unstable disk. *Nature* **538**, 483-486 (2016).
3. Wurster, J. & Bate, M.R. Disc formation and fragmentation using radiative non-ideal magnetohydrodynamics. *Mon. Not. R. Astron. Soc.* **486**, 2587-2603 (2019).
4. Tobin, J.J. et al. The VLA/ALMA Nascent Disk and Multiplicity (VANDAM) Survey of Perseus Protostars. VI. Characterizing the Formation Mechanism for Close Multiple Systems. *Astrophys. J.* **867**, 43 (2018).
5. Zhao, B. Caselli, P., Li, Z.-Y. & Krasnopolsky, R. Decoupling of magnetic fields in collapsing protostellar envelopes and disc formation and fragmentation. *Mon. Not. R. Astron. Soc.* **473**, 4868-4889 (2018).
6. Sigalotti, L.D.G., Cruz, F., Gabbasov, R., Klapp, J. & Ramírez-Velasquez, J. From Large-scale to Protostellar Disk Fragmentation into Close Binary Stars. *Astrophys. J.* **857**, 40 (2018).
7. Takakuwa, S. et al. Spiral Arms, Infall, and Misalignment of the Circumbinary Disk from the Circumstellar Disks in the Protostellar Binary System L1551 NE. *Astrophys. J.*, **837**, 86 (2017).

8. Matsumoto, T., Saigo, K., & Takakuwa, S. Structure of a Protobinary System: An Asymmetric Circumbinary Disk and Spiral Arms. *Astrophys. J.*, **871**, 36 (2019).
9. Thun, D., Kley, W., & Picogna, G. Circumbinary discs: Numerical and physical behaviour. *Astron. Astrophys.,* **604**, 102 (2017).
10. Mösta, P., Taam, R.E., & Duffell, P.C. Gas Flows Within Cavities of Circumbinary Disks in Eccentric Binary Protostellar Systems. *Astrophys. J. Lett.*, **875**, 21 (2019).
11. Le Gouellec, V. J. M. et al. Characterizing magnetic field morphologies in three Serpens protostellar cores with ALMA. *Astrophys. J.*, **885**, 106 (2019).
12. Hull, C. L. H., Le Gouellec, V. J. M., Girart, J. M., Tobin, J. J., Bourke, Tyler. L. Understanding the origin of the magnetic field morphology in the wide-binary protostellar system BHR 71. *Astrophys. J.* **892**, 151 (2020).
13. Kuffmeier, M., Haugbølle, T. & Nordlund, Å. Zoom-in Simulations of Protoplanetary Disks Starting from GMC Scales. *Astrophys. J.* **846**, 7 (2017).
14. Kuffmeier, M., Frimann, S., Jensen, S.S., & Haugbølle, T., Å. Episodic accretion: the interplay of infall and disc instabilities. *Mon. Not. R. Astron. Soc.* **475**, 2642-2658 (2018).
15. Andrews et al. DSHARP I. Motivation, Sample, Calibration, and Overview. *Astrophys. J.* Lett., **869**, 41 (2018).
16. Huang et al. DSHARP II. Characteristics of Annular Substructures. *Astrophys. J.* Lett., **869**, 42 (2018).
17. Zucker, Catherine et al. Mapping Distances across the Perseus Molecular Cloud Using CO Observations, Stellar Photometry, and Gaia DR2 Parallax Measurements. *Astrophys. J.* **869**, 83 (2018).
18. Stephens, I.W. et al. Mass Assembly of Stellar Systems and Their Evolution with the SMA (MASSES)—Full Data Release *Astrophys. J. Supplement Series*, **245**, 21 (2019).
19. Frimann, S et al. Protostellar accretion traced with chemistry. High-resolution $C^{18}O$ and continuum observations towards deeply embedded protostars in Perseus. *Astron. Astrophys.,* **602**, 120 (2017).
20. Hsieh, T.-H. et al. Chronology of Episodic Accretion in Protostars -- an ALMA survey of the CO and $H_2O$ snowlines. *Astrophys. J.* **884**, 149 (2019).
21. Shu, Li & Allen. Does Magnetic Levitation or Suspension Define the Masses of Forming Stars? *Astrophys. J.*, **601**, 930-951, (2004).
22. Dunham et al. The Evolution of Protostars: Insights from Ten Years of Infrared Surveys with Spitzer and Herschel. Protostars and Planets VI, Henrik Beuther, Ralf S. Klessen, Cornelis P. Dullemond, and Thomas Henning (eds.), University of Arizona Press, Tucson, 914 pp., p. 195-218, (2014)
23. Wurster, J. & Li, Z.-Y. The role of magnetic fields in the formation of protostellar discs. *Frontiers in Astronomy and Space Sciences*, **5**, 39 (2018).
24. Kuznetsova, A., Hartmann, L. & Heitsch, F. The Origins of Protostellar Core Angular Momenta *Astrophys. J.* **876**, 33 (2019).
25. Kuffmeier, M., Calcutt, H. & Kristensen, L. E The bridge: a transient phenomenon of forming stellar multiples. Sequential formation of stellar companions in filaments around young protostars. *Astron. Astrophys.,* **628**, 112 (2019).
26. Wurster, J., Bate, M. R., & Price, D. J. There is no magnetic braking catastrophe: low-mass star cluster and protostellar disc formation with non-ideal magnetohydrodynamics. *Mon. Not. R. Astron. Soc.* **489**, 1719-1741 (2019).
27. Yen, H.-W. et al. ALMA Observations of Infalling Flows toward the Keplerian Disk around the Class I Protostar L1489 IRS. *Astrophys. J.* **793**, 1 (2014).

28. Yen, H.-W. et al. HL Tau Disk in HCO$^+$ (3-2) and (1-0) with ALMA: Gas Density, Temperature, Gap, and One-arm Spiral. *Astrophys. J.* **880**, 2 (2019).
29. Mendoza, S., Tejeda, E. & Nagel E. Analytic solutions to the accretion of a rotating finite cloud towards a central object – I. Newtonian approach. *Mon. Not. R. Astron. Soc.* **393**, 579-586 (2009).
30. van der Tak, F. F. S.; Black, J. H.; Schöier, F. L.; Jansen, D. J.; & van Dishoeck, E. F. A computer program for fast non-LTE analysis of interstellar line spectra. With diagnostic plots to interpret observed line intensity ratios. *Astron. Astrophys.,* **468**, 627-635 (2007).
31. Evans, N. J. II et al. Detection of Infall in the Protostar B335 with ALMA. *Astrophys. J.,* **814**, 22 (2015).
32. Lee, C.-F., Li, Z.-Y., & Turner, N. J., Spiral structures in an embedded protostellar disk driven by envelope accretion. Nat. Astron. 466 (2019).
33. Ellithorpe, E. A., Duchene, G. & Stahler, S. W. The Nature of Class I Sources: Periodic Variables in Orion. *Astrophys. J.* **885**, 64 (2019).
34. Bergin, E.A. & Tafalla, M. Cold Dark Clouds: The Initial Conditions for Star Formation. *Annu. Rev. Astron. Astrophys.*, **45**, 339-396 (2007).
35. Sakai, N., & Yamamoto, S., *Chem. Rev.*, **113**, 8981 (2013).
36. Nanne, J. A. M., Nimmo, F., Cuzzi, J. N. & Kleine, T. *Origin of the non-carbonaceous-carbonaceous meteorite dichotomy*, Earth and Planetary Science Letters, **511**, 44-54, (2019).

**Acknowledgements**
JEP thanks Andreas Burkert, Alyssa A. Goodman, Stella S.R. Offner, and Ralf S. Klessen for discussions and comments that improved the paper.

Based on observations carried out with the IRAM Interferometer NOEMA. IRAM is supported by INSU/CNRS (France), MPG (Germany) and IGN (Spain). This paper makes use of the following ALMA data: ADS/JAO.ALMA\#2013.1.00031.S. ALMA is a partnership of ESO (representing its member states), NSF (USA) and NINS (Japan), together with NRC (Canada), MOST and ASIAA (Taiwan), and KASI (Republic of Korea), in cooperation with the Republic of Chile. The Joint ALMA Observatory is operated by ESO, AUI/NRAO and NAOJ.

This research made use of Astropy, a community-developed core Python package for Astronomy[55], and APLpy, an open-source plotting package for Python hosted at *http://aplpy.github.com*.

**Author Contributions**
JEP led the project and reduced the ALMA data, led the project and imaged the NOEMA data. NC and RN reduced the NOEMA data. JEP wrote the manuscript. All authors contributed to the NOEMA proposal, discussed the results and implications, and commented on the manuscript.

**Competing financial interests**
The authors declare no competing financial interests.

**Corresponding author**
Correspondence and requests for materials should be addressed to Jaime E. Pineda (jpineda@mpe.mpg.de).

# Figures

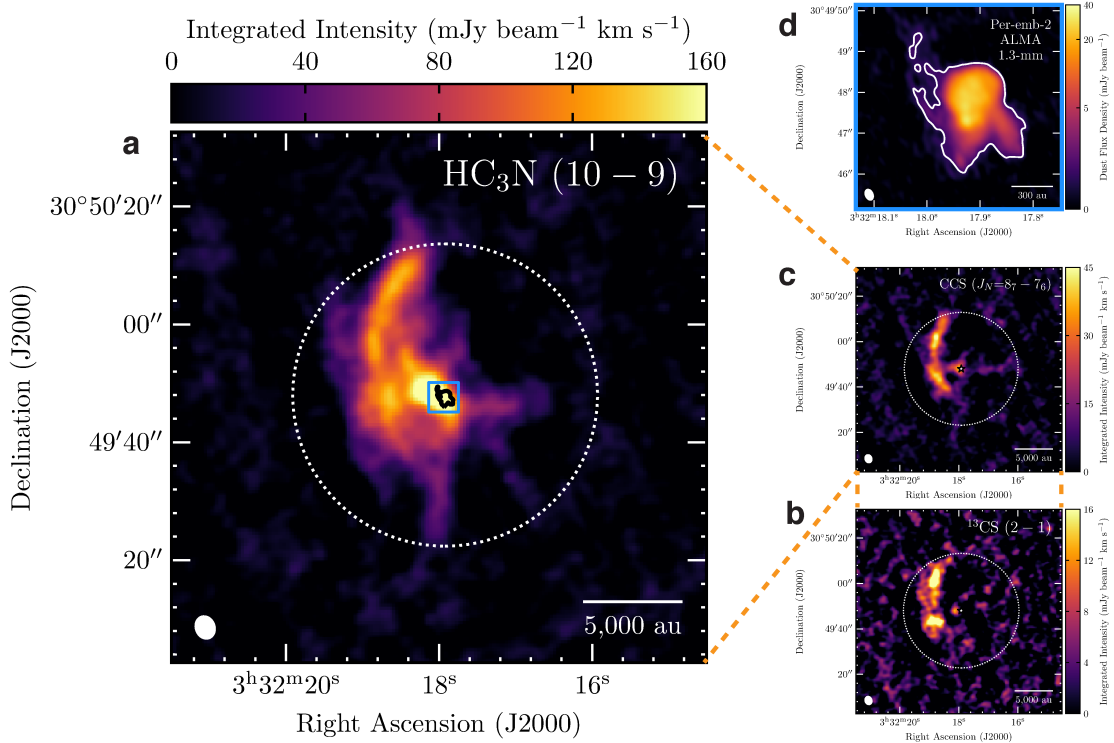

**Figure 1** Image of the "chemically fresh" material surrounding the protostar (panels a, b, and c) and zoomed-in high-angular resolution image of the material near disks scales (panel d). The panels a, b, and c reveal a streamer of fresh material starting at a distance on the sky of 10,500 au from the central protostar. The streamer is seen in two carbon-chain molecules and a S-bearing species, all known to be abundant in chemically young material. The black contour in panel a corresponds to the asymmetrical and feather-like feature near disk scales, and it is also shown as a white contour in the zoomed-in image in the panel d. The angular resolution of the observations and scalebar are shown in the bottom left and bottom right corner, respectively. The 50% primary beam response is shown by the white dotted circle. The ALMA 1.3-mm contour levels are shown in panels a and d at 1mJy beam$^{-1}$, where the noise level is 0.166 mJy beam$^{-1}$.

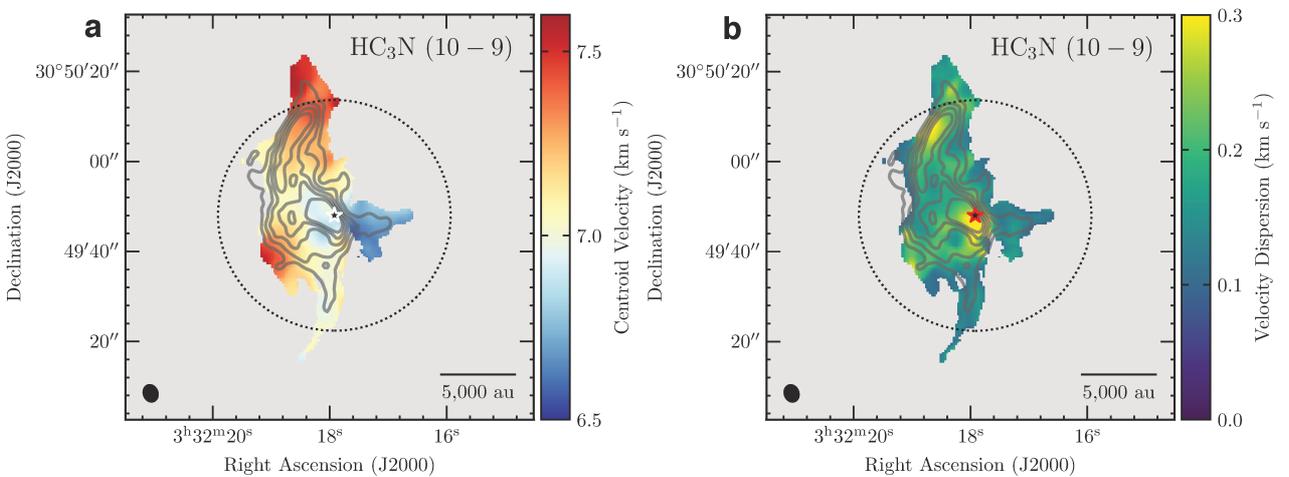

**Figure 2** Centroid velocity and velocity dispersion of the streamer tracing HC$_3$N (10-9) emission in panels a and b, respectively. The velocity field is smooth and shows a clear velocity gradient. The velocity dispersion shows subsonic levels of turbulence, and therefore the material is not affected

by the outflow. The angular resolution of the observations and scalebar are shown in the bottom left and bottom right corner, respectively. The 50% primary beam response is shown by the black dotted circle. The HC$_3$N contours levels are drawn at 5, 8, 11, 14 and 17x rms, where rms is 8 mJy beam$^{-1}$ km s$^{-1}$.

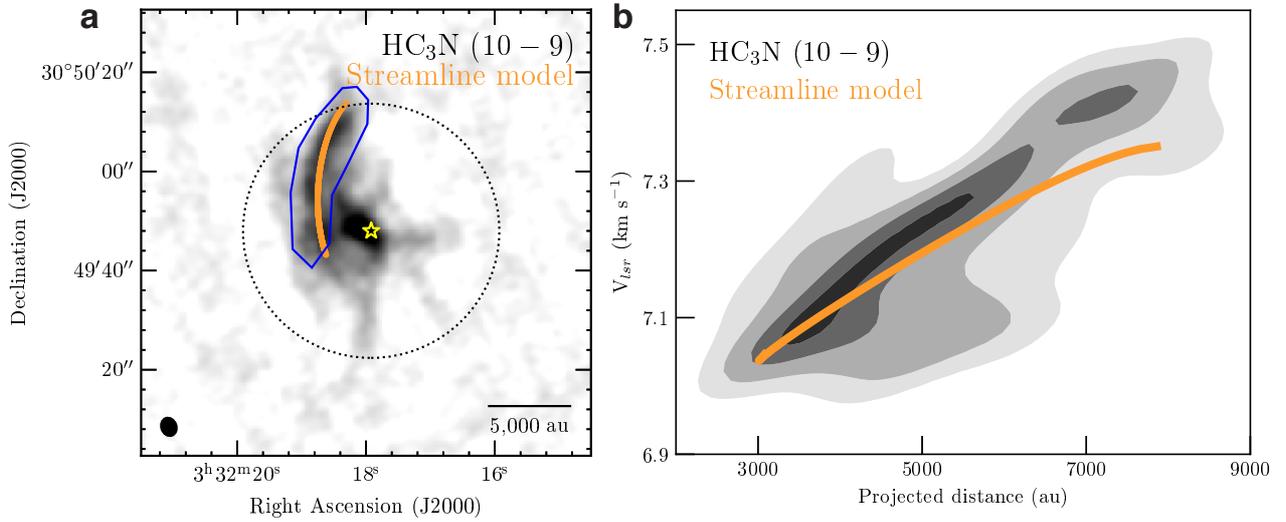

**Figure 3** Sky position and velocity along the line-of-sight of the streamline model (in orange) overlaid on the observations (gray). The streamer is highlighted by the blue polygon in panel a. The panel b shows varying levels of kernel density estimation of the velocity as a function of projected distance from the central system. The contour levels are drawn starting at 0.5σ and progressing outwards in steps of 0.5σ, where the σ-levels are equivalent to that of a bivariate normal distribution. The streamline model provides a good match to the observations and it supports the interpretation of gravitational free-fall of the streamer towards the central regions of the dense core. The angular resolution of the observations and scalebar are shown in the bottom left and bottom right corner, respectively. The 50% primary beam response is shown by the black dotted circle.

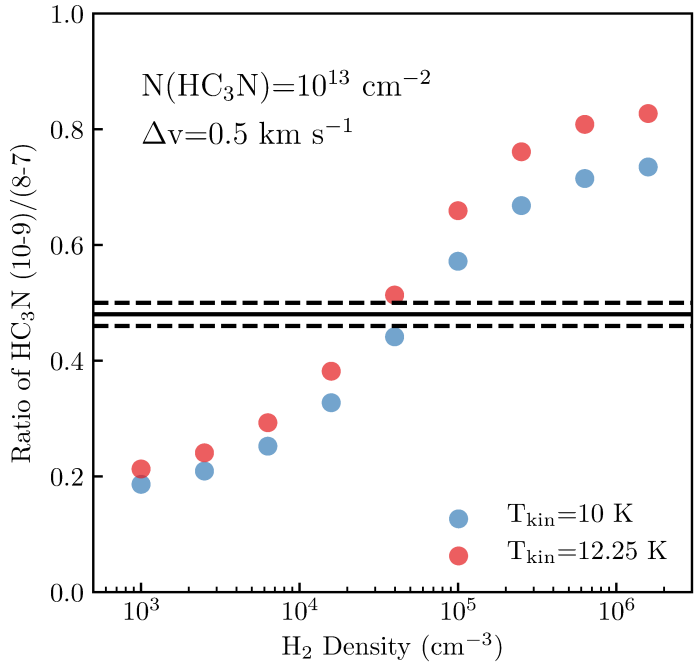

**Extended Data Figure 1:** Ratio of the integrated intensity of HC$_3$N (10-9) to (8-7) transitions as a function of H$_2$ density. The ratio is calculated using RADEX, for kinetic temperatures of 10 and 12.25 K in blue and red, respectively. The measured average value and associated uncertainty of the ratio in the streamer is marked by the horizontal solid and dashed lines, respectively. This comparison shows that the average density in the streamer is (4±2)x10$^4$ cm$^{-3}$.

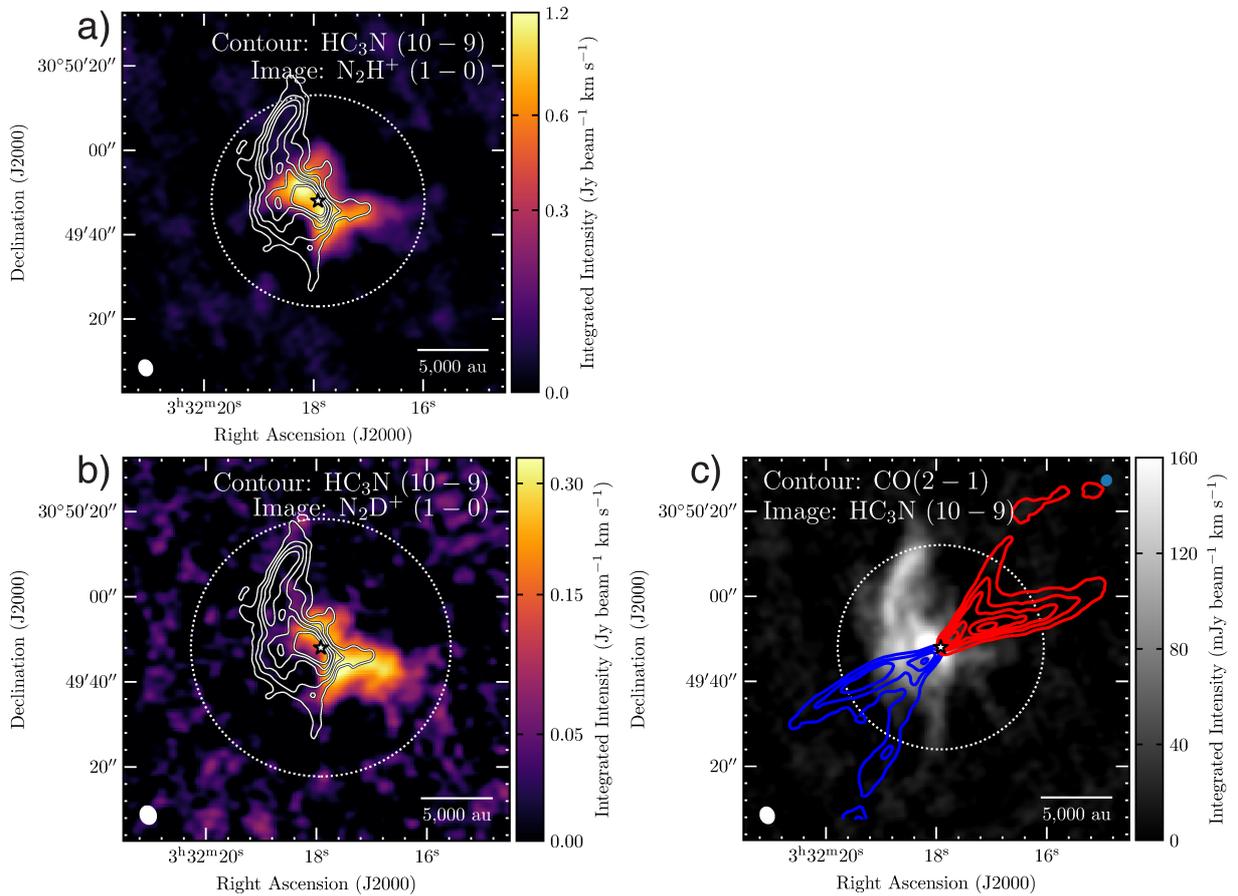

**Extended Data Figure 2:** The streamer is unrelated to the old inner (radius <3,000 au) envelope or to the outflow emission from the young protostellar system. Comparison between HC$_3$N emission tracing the streamer and "chemically old" dense gas in panels a and b, and with the outflow in panel c. Panels a and b show the dense gas traced by N$_2$H$^+$ and N$_2$D$^+$ in the background, with the contours of the chemically fresh material. The streamer is clearly mostly outside the N$_2$H$^+$ and N$_2$D$^+$ emission, and it shows a significantly different morphology to the inner envelope. Panel c shows the outflow emission traced by CO (2-1) in red and blue contours, while the background emission is the HC$_3$N integrated intensity. This shows that the streamer is unrelated to the outflow interaction. The 50% primary beam response for the $N_2H^+$, $N_2D^+$, and CO (2-1) observations are shown by the white dotted circle in the left and right panels, respectively. The $HC_3N$ contours levels are drawn at 5, 8, 11, 14 and 17x rms, where rms is 8 mJy beam$^{-1}$ km s$^{-1}$. The CO contours levels are drawn at 5, 15, 25, 35 and 45x rms, where rms is 0.2 Jy beam$^{-1}$ km s$^{-1}$.

**Methods**

**Observations and data reduction: IRAM NOrthern Extended Millimeter Array (NOEMA)**

We conducted NOEMA observations of the Per-emb-2 region on 2018 June 22, July 27-28, and August 19 in the D configuration, while on 2018 October 9, 10, 17, and 18 in the C configuration (project S18AG), with baselines between 16.4 and 287.8 meters. This corresponds to a maximum recoverable scale of 30 arcsec. We used the Band 1 receiver and configured the PolyFix correlator with a LO frequency of 82.505 GHz and an instantaneous bandwidth of 31 GHz spread over two sidebands (upper and lower) and two polarisations and with a spectral resolution of 2 MHz. Additionally, high-resolution chunks are placed with a width of 64 MHz and a fixed spectral resolution of 62.5 kHz targeting several molecular transitions, including $HC_3N$ (J=10-9) and (J=8-7), CCS ($J_N=8_7-7_6$), $^{13}CS$ (2-1), $N_2H^+$(1-0), and $N_2D^+$ (J=1-0). The integrated intensity map of the $HC_3N$ (J=10-9), CCS ($J_N=8_7-7_6$), $^{13}CS$ (2-1) lines are shown in Figure 1.

The data are reduced and calibrated using standard observatory pipeline in CLIC, which is part of the GILDAS software package (http://www.iram.fr/IRAMFR/GILDAS). The bandpass calibrators used are 3C84 (4 tracks) and 3C454.3 (3 tracks). Phase and amplitude calibration was performed every 20 minutes using J0333+321 and J0329+351 for all observations. The absolute flux calibration was performed using the standard flux calibrators MWC349 and LkHa101 and the flux calibration uncertainty is within 15%.

Continuum subtraction and data imaging were performed using the GILDAS software version from March 2019 (mar19b). We image the continuum emission of each baseband (LI, LO, UI, and UO) using the line free channels in each baseband with the *uv_continuum* command. The continuum emission is strong enough to allow for a self-calibration solutions down to 45 seconds integration for the phase of the continuum using the *selfcal* command. We apply these solutions to the continuum data as well as the corresponding line data using the task *uv_cal*. The line emission is imaged using the continuum subtracted data generated with the *uv_baseline* command and a first order baseline.

The final noise level (measured using the task *noise* in GILDAS) in the $HC_3N$ (J=10-9), $HC_3N$ (J=8-7), CCS ($J_N=8_7-7_6$), $^{13}CS$ (2-1), $N_2D^+$ (J=1-0), and $N_2H^+$ (J=1-0) cubes are 5.5, 8.9, 6.3, 6.3, 6.6, and 6.2 mJy beam$^{-1}$ channel$^{-1}$, respectively, while the integrated intensity maps have noise levels of 8, 10, 4, 3, 19, and 30 mJy beam$^{-1}$ km s$^{-1}$, respectively, and the beam sizes are 3.7x3.0 (19deg), 4.4x3.7(19deg), 3.6x2.9(19deg), 3.7x3.0(19deg), 4.2x3.5(21deg), and 3.6x3.0(20deg), respectively.

**Observations and data reduction: Atacama Large Millimeter Array (ALMA)**

Per-emb-2 was observed with the 12-m ALMA array using the Band 6 receivers on 2015 September 27 under project 2013.1.00031.S. The array configuration included 34 antennas with baselines between 43.3 and 2,270 m and a total on-source time of 3.5 min. The standard pipeline calibration was done using Common Astronomy Software Applications (CASA) package 4.5.0[37], while imaging was done in CASA 5.4.1. The 1.3 mm continuum image was obtained using the *tclean* task from single broad band continuum spectral window of the configuration and imaged using multiscale (with scales of 0, 0.24, and 0.72 arcsec) with a robust Briggs parameter of 0.5 to achieve an angular resolution of 0.275 arcsec x 0.185 arcsec (PA=23.6º), while dropping all baselines shorter than 50 kλ to reduce imaging artifacts due to resolved out emission as done in the original paper presenting the data[4]. We perform self-calibration both in phase and amplitude down to a single integration (6.05 seconds) and to 1200 seconds, respectively. The final self-calibrated image was improved substantially, both by removing image artifacts and reducing the noise level. The final image is shown in panel d of Figure 1, and it achieved an rms noise of 166 μJy beam$^{-1}$, which is similar to the previous results[4], although with a better recovery of some of the asymmetric features. Polarisation observations of the continuum emission at 870μm using ALMA also show similar features at slightly lower angular resolution[38].

**Physical properties of the Streamer**

We calculate the ratio between the HC$_3$N (10-9) and (8-7) integrated intensity maps, after they are smoothed and re-gridded to a common resolution and grid. The ratio of two transitions of HC$_3$N is a good probe of the density[39]. We measure the mean value of this ratio in the streamer to be 0.48±0.02, where the correction for the primary beam response is applied. We also compute this ratio using RADEX[30], with a column density of HC$_3$N of $10^{13}$ cm$^{-2}$, a line-width of 0.5 km s$^{-1}$ (consistent with the observations), gas kinetic temperatures of 10 and 12.25 K (consistent with the average core temperature of 11.7 K[40]), and densities from $10^3$ up to $2 \times 10^6$ cm$^{-3}$. The comparison between the measurement and the RADEX models are shown in Extended Data Figure 1, with the average value of the ratio in the streamer marked by the horizontal black line. The comparison clearly shows that a density of $(4\pm2) \times 10^4$ cm$^{-3}$ reproduces the observations, and it is our estimated average density within the polygon marking the streamer.

We determine the column density in the optically thin regime,

$$N_{(J+1)} = \frac{8\pi k \nu^2}{h c^3 A_{J(J+1)}} \int T dV ,$$

for the level (J+1)=10. Then we use the conversion factor between N(HC$_3$N)$_{(J=10)}$ into N(HC$_3$N)$_{total}$ for the best RADEX model of 0.01963, and derive an average HC$_3$N total column density of $2.0 \times 10^{13}$ cm$^{-2}$ in the streamer.

The total mass is finally determined by using an [HC$_3$N/H$_2$] abundance ratio, and then calculating the total mass in H$_2$. The abundance is expected to lie between those found toward the chemically young carbon rich peak of TMC-1 ([HC$_3$N/H$_2$]$_{TMC1-CP}$=2.8×10$^{-9}$)[41] and the more evolved L134N ([HC$_3$N/H$_2$]$_{L134N}$=3×10$^{-10}$)[42]. Therefore, the total mass in the streamer is between 0.1 and 1 M$_{sun}$.

We combine the streamer's mean H$_2$ density, mean total HC$_3$N column density, and the typical observed width of 15 arcsec (4,500 au) to estimate the streamer's [HC$_3$N/H$_2$]$_{Streamer}$ abundance as

$$\left[ \frac{\mathrm{HC_3N}}{\mathrm{H_2}} \right]_{Streamer} = \frac{N(\mathrm{HC_3N})}{n_{\mathrm{H_2}} \, width} = 7.4 \times 10^{-9} .$$

This abundance in the streamer is comparable to that of the carbon rich peak of TMC-1, and therefore consistent with a chemically fresh streamer. Therefore, we also adopt a conservative estimate of the streamer mass of 0.1 M$_{Sun}$.

**Line fit**

We fit single gaussian models to each pixel of the HC$_3$N (8-7) and (10-9) data cubes, using the *PySpecKit* Python package[43], and a rest frequency (obtained from CDMS, https://cdms.astro.uni-koeln.de/) of 72783.822 and 90979.023 MHz, respectively. In addition, we also fit the N$_2$D$^+$ (1-0) lines using all the hyperfine components using also *PySpecKit*.

The centroid velocity and velocity dispersion maps derived from HC$_3$N (10-9) are shown in Figure 2, and with the corresponding integrated intensity shown in contours. The observed velocity dispersion is comparable to a level of turbulence below 0.5× the sound speed at 10 K ($\mathcal{M}_s = \sigma_{turb}/c_s < 0.5$), similar to those seen in dense cores[44,45]. This velocity dispersion is consistent with fresh material unaffected by the molecular outflow driven by the young protostar.

**Is the streamer in free-fall collapse?**

In the case of Pressure-Free or Free-Fall Collapse it is possible to estimate both the speed and the time scale of collapse[46] for a given mass and radius. We estimate the free fall speed as

$$v_{ff} = -\sqrt{\frac{2GM}{R}},$$

where $M$ is the enclosed mass at radius $R$, the free-fall timescale as

$$t_{ff} = \sqrt{\frac{R^3}{GM}},$$

and we also estimate the velocity gradient as

$$\mathrm{grad}_{v,ff} = \sqrt{\frac{GM}{2R^3}}.$$

The streamer is detected starting (farthest from the protostar) at a distance of 35 arcsec (10,500 au at the distance of Perseus) from the protostar. The previously estimated total dense core mass is 3.2 M$_{sun}$[47] and a deconvolved radius of 4,400 au[48], after correcting for the updated distance to Perseus, which also corresponds to an average dense core density of 1.4x10$^6$ cm$^{-3}$. Using these values we estimate the values in free-fall collapse at the beginning (farthest from the protostar) of the streamer: $v_{ff}$ =1.1 km s$^{-1}$, $t_{ff}$ = 96 kyr, and grad$_{v,ff}$ = 7 km s$^{-1}$ pc$^{-1}$.

We use the HC$_3$N (10-9) velocity map to estimate the streamer kinematical properties. There is a smooth velocity gradient in the streamer (7.4 km s$^{-1}$ - 6.98 km s$^{-1}$) at a velocity similar from the gas around the young protostar (7.05 km s$^{-1}$), and stretching over an extent of 35 arcsec (10,500 au). This corresponds to a velocity gradient of 11.5 km s$^{-1}$ pc$^{-1}$. This is within 60% from the free-fall acceleration estimated above.

**Calculation of Streamlines**

We model the streamline using the analytic solutions for material moving from a rotating finite cloud towards a central object[29]. This approach provides a solution along a given streamline which also provides a velocity along the line-of-sight, and it depends on the central mass (M$_{center}$), initial position and velocity of the streamer in spherical coordinates (r$_0$, θ$_0$, φ$_0$, v$_{r0}$, v$_{θ0}$). The implemented solution for the streamline is calculated with a mass of 3.2 M$_{Sun}$ (equals to the dense core mass), initial distance of 9,000 au from the central object, initial tangential velocity 0.53 km s$^{-1}$ (consistent with the difference in V$_{lsr}$ seen between the tip of the steamer and the protostar), and without an initial radial velocity. We apply two rotations due to inclination angle and the position angle to directly compare to the observations. The angles $\theta_0$ and $\varphi_0$ are found by matching the initial position of the streamline with the seen extent and position of the streamer.

Left panel of Figure 3 shows the projected trajectory of the streamline (in red), which matches well the streamer seen in integrated intensity (background image). The right panel of Figure 3 shows the velocity along the line of sight for the streamline in red as a function of projected separation from the central system, and the background is the kernel density estimate (KDE) of the velocity and projected separation measured in the data. The KDE allow us to show the overall properties of the streamer and its variation, without the need for defining a particular cut. We use the KDE implementation in *scipy* and the KDE bandwidth selection following Scott's rule[49]. Both comparisons show a good match with the data, with only small disagreements between the velocity along the line of sight of the model and data. The model requires a faster rotation (more specific angular momentum) to match the velocity at large radii, and a slower rotation at smaller radii (less specific angular momentum). This suggests that a larger amount of specific angular momentum is required in the outer regions than in the inner regions. This might be due to the loss of angular momentum, as previously seen in other protostellar dense cores[50]. In addition, this simple infall model does not take into account the protostellar mass, or the variation of the enclosed mass as a function of radius, which could also generate small variations to the streamer trajectory and velocity.

**Protostellar accretion rate**

The protostellar accretion rate is determined by modelling the observed bolometric luminosity, $L_{bol}$, as dominated by the accretion rate on the central protostellar object as

$$L_{bol} = \frac{G M_* \dot{M}_*}{R_*},$$

where $M_*$ is the protostellar mass, $R_*$ is the protostellar radius, and $\dot{M}_*$ is the accretion rate. We estimate the accretion rate using a 0.25 $M_{Sun}$ protostellar mass (close to peak of the IMF), a 3 $R_{Sun}$ stellar radius, and $L_{bol}$=1.9 $L_{Sun}$ for an accretion rate of 7x10$^{-7}$ $M_{Sun}$ yr$^{-1}$.

The value of the protostellar mass is not well constrained due to the lack of Keplerian rotation measured, but using a brown dwarf mass (0.076 $M_{Sun}$) provides a strong upper limit to the accretion rate of 2x10$^{-6}$ $M_{Sun}$ yr$^{-1}$. This is in the upper end of accretion rates estimated toward Class 0 protostars[47].

**Depletion of Carbon species from the gas phase**

The streamer is dominated by carbon-chain species, however, these species are typically depleted in dense cores[51]. The "freeze-out" timescale is estimated as

$$t_{dep} = \frac{8,000}{(S\, n_5)} \text{ yr,}$$

with $n_5$ is the H$_2$ number density in units of 10$^5$ cm$^{-3}$ and $S$ the sticking probability (of order unity[52,53]) for CO on grains. This estimate is appropriate for the species studied here (carbon-chain molecules and an early-type molecule): $^{13}$CS, CCS, and HC$_3$N[34]. Using the average dense core density (1.4x10$^6$ cm$^{-3}$), we estimate the depletion timescale in the dense core of 600 yr, therefore, these species are not expected in the 'old' dense core. However, the average density in the streamer (over a 10,500 au region) derived using RADEX implies a timescale of 20,000 yrs, which is short compared to the Class 0 lifetime of ~100,000 yrs. This encourage the use of the term 'chemically fresh material'.

**On the outflow orientation**

We trace the outflow emission with the CO (2-1) data from the MASSES survey[18] with a 2.7"x2.5" beam obtained from robust=1 parameter. The blue and red emission is calculated between [-1.5, 5] km s$^{-1}$ and [9, 16] km s$^{-1}$ and has an rms of 0.2 Jy beam$^{-1}$ km s$^{-1}$ for both maps. These red and blue lobes are shown in Extended Data Figure 2.

The 10 arcmin scale of the outflow, as traced with $H_2$ images and connected to some HH objects, show that no recent event of a substantial change in the outflow orientation has occurred[54]. This rules out the possibility that a previous different outflow orientation (e.g. due to misaligned accretion events) could be responsible for the release or excitation of the carbon species from the grain.

**Data Availability**

The data and analysis that support the findings of this study are available in GitHub (repository https://github.com/jpinedaf/NOEMA_streamer_analysis) with the identifier DOI 10.5281/zenodo.3874992.

**References**

37. McMullin, J. P.; Waters, B.; Schiebel, D.; Young, W. & Golap, K. CASA Architecture and Applications, Astronomical Data Analysis Software and Systems XVI ASP Conference Series, Vol. 376, proceedings of the conference held 15-18 October 2006 in Tucson, Arizona, USA. Edited by Richard A. Shaw, Frank Hill and David J. Bell., p.127 (2007).
38. Cox, E. G. et al. ALMA's Polarized View of 10 Protostars in the Perseus Molecular Cloud. *Astrophys. J.* **855**, 92 (2018).
39. Li, Di & Goldsmith, Paul F. Is the Taurus B213 Region a True Filament?: Observations of Multiple Cyanoacetylene Transitions. *Astrophys. J.* **756**, 165-169 (2012).
40. Rosolowsky, E. W. et al. An Ammonia Spectral Atlas of Dense Cores in Perseus. *Astrophys. J. Supplement Series*, **175**, 509-521 (2008).
41. Loomis, Ryan A. et al. Non-detection of HC11N towards TMC-1: constraining the chemistry of large carbon-chain molecules. *Mon. Not. R. Astron. Soc.* **463**, 4175-4183 (2016).
42. Dickens, J. E. et al. A Study of the Physics and Chemistry of L134N. *Astrophys. J. 542,* 870-889 (2000).
43. Ginsburg, Adam & Mirocha, Jordan. PySpecKit: Python Spectroscopic Toolkit. Astrophysics Source Code Library, record ascl:1109.001 (2011).
44. Pineda, J. E. et al. Direct Observation of a Sharp Transition to Coherence in Dense Cores. *Astrophys. J. Lett.*, **712**, L116-L121 (2010).
45. Pineda, J. E. et al. The formation of a quadruple star system with wide separation. *Nature* **518**, 213-215 (2015).
46. Stahler, Steven W. & Palla, Francesco. The Formation of Stars. The Formation of Stars, by Steven W. Stahler, Francesco Palla, pp. 865. ISBN 3-527-40559-3. Wiley-VCH , January (2005).
47. Mottram, J. C. et al. Outflows, infall and evolution of a sample of embedded low-mass protostars. The William Herschel Line Legacy (WILL) survey. *Astron. Astrophys.,* **600**, 99 (2017).

48. Enoch, M. L. et al. Bolocam Survey for 1.1 mm Dust Continuum Emission in the c2d Legacy Clouds. I. Perseus. *Astrophys. J. Lett.*, **638**, 293-313 (2006).
49. Scott, D.W. "Multivariate Density Estimation: Theory, Practice, and Visualization", John Wiley & Sons, New York, Chicester, (1992).
50. Pineda, J. E. et al. The Specific Angular Momentum Radial Profile in Dense Cores: Improved Initial Conditions for Disk Formation. *Astrophys. J.*, **882**, 103 (2019).
51. Caselli, P., Walmsley, C.M., Tafalla, M., Dore, L., Myers, P.C. CO Depletion in the Starless Cloud Core L1544. *Astrophys. J. Lett.* **523**, 165-169 (1999).
52. Jones, A. P. & Williams, D. A. Time-dependent sticking coefficients and mantle growth on interstellar grains. *Mon. Not. R. Astron. Soc.*, **217**, 413-421 (1985).
53. Tielens, A. G. G. M., & Allamandola, L. J. Composition, Structure, and Chemistry of Interstellar Dust. Interstellar Processes, Proceedings of a symposium, held at Grand Teton National Park, Wyo., July, 1986, Dordrecht: Reidel, edited by David J. Hollenbach, and Harley A. Thronson. Astrophysics and Space Science Library, **134**, 397 (1987).
54. Walawender, J., Reipurth, B. & Bally, J. Multiple Outflows and Protostars in Barnard 1. II. Deep Optical and Near-Infrared Images. *Astron. J.*, **137**, 3254-3262 (2009).
55. Astropy Collaboration et al. Astropy: A community Python package for astronomy. Astron. Astrophys. 558, A33 (2013).